\documentclass[a4paper]{article}
\pdfoutput=1

\usepackage[english]{babel}
\usepackage[utf8x]{inputenc}
\usepackage[T1]{fontenc}
\usepackage{authblk}

\usepackage[top=2cm,bottom=2cm,left=3cm,right=3cm,marginparwidth=2cm]{geometry}

\usepackage{amsfonts}
\usepackage{amsmath}
\usepackage{graphicx}
\usepackage[colorinlistoftodos]{todonotes}
\usepackage[colorlinks=true, allcolors=blue]{hyperref}

\title{A Graph Gaussian Embedding Method for Predicting Alzheimer’s Disease Progression with MEG Brain Networks}
\date{}
\author[1]{Mengjia Xu}
\author[2]{David Lopez Sanz}
\author[2]{Pilar Garces}
\author[2]{Fernando Maestu}
\author[3*]{Quanzheng Li}
\author[1*]{Dimitrios Pantazis}

\affil[1]{McGovern Institute for Brain Research, Massachusetts Institute of Technology, Cambridge, MA 02139, USA}
\affil[2]{Department of Experimental Psychology, Complutense University of Madrid, Madrid 28040, Spain}
\affil[3]{Department of Radiology, Harvard Medical School, Boston, MA 02114, USA}
\affil[*]{Corresponding author. Email: pantazis@mit.edu; li.quanzheng@mgh.harvard.edu}

\usepackage{amsmath}
\begin{document}
\maketitle
\begin{abstract}
Characterizing the subtle changes of functional brain networks associated with the pathological cascade of Alzheimer’s disease (AD) is important for early diagnosis and prediction of disease progression prior to clinical symptoms. We developed a new deep learning method, termed multiple graph Gaussian embedding model (MG2G), which can learn highly informative network features by mapping high-dimensional resting-state brain networks into a low-dimensional latent space. These latent distribution-based embeddings enable a quantitative characterization of subtle and heterogeneous brain connectivity patterns at different regions, and can be used as input to traditional classifiers for various downstream graph analytic tasks, such as AD early stage prediction, and statistical evaluation of between-group significant alterations across brain regions. We used MG2G to detect the intrinsic latent dimensionality of MEG brain networks, predict the progression of patients with mild cognitive impairment (MCI) to AD, and identify brain regions with network alterations related to MCI.
\end{abstract}

\section*{Introduction}
Alzheimer’s disease (AD) is a progressive neurodegenerative disorder and the most common type of dementia \cite{AD2016lancet}. It is characterized by the progressive destruction of axonal pathways, which leads to aberrant network changes at both anatomical and functional levels \cite{lynn2019physics}. Mild cognitive impairment (MCI) is a key prodromal stage of AD and has a heterogeneous evolution pattern, i.e., 10-15\% MCI patients may progress to AD or other types of dementia per year, whereas the others may remain stable or improve over time \cite{davatzikos2011prediction}. Thus, there is an urgent need to effectively identify progressive MCI (pMCI) and stable MCI (sMCI) populations during the early stages of AD. This could lead to a better understanding of the underlying heterogeneous pathogenesis of AD, provide insights into the design of new clinical trials, and inform mechanistic-based therapeutic interventions. 

The first set of clinical criteria for MCI conversion prediction and diagnosis of AD focused on symptoms and neurophysiological tests, such as the clinical dementia rating-CDR, mini-mental state examination (MMSE) score, and Montreal Cognitive Assessment (MoCA) score \cite{hoops2009validity}. However, the advent of biomarkers that measure pathological changes in vivo has provided new tools to predict and understand the complex etiology of AD. The design of AD biomarkers has been largely based on quantification of A$\beta$ and Tau protein deposition levels, and assessment of neurodegeneration \cite{jack_atn:_2016}. A$\beta$ protein accumulation levels are measured by amyloid PET \cite{klunk2004imaging} or low cerebrospinal fluid (CSF) A$\beta_{42}$ \cite{fagan2007cerebrospinal}. Tau protein accumulation levels are measured by tau PET \cite{mattsson2009csf} or elevated CSF phosphorylated tau (p-tau) \cite{buerger2006csf}. Neurodegeneration is assessed by CSF total tau (t-tau) \cite{besson2015cognitive}, $^{18}$F- fluorodeoxyglucose positron emission tomography (FDG-PET) \cite{besson2015cognitive}, and atrophy on structural MRI \cite{t2020association}. Additional information can be obtained by AD genetic risk indicators, such as  apolipoprotein E (ApoE), which is coded by the APOE gene \cite{kunkle2019genetic, verghese2011apolipoprotein}.

Despite these tremendous strides in biomarker technology, MRI and PET biomarkers are expensive and PET scans involve exposure to radioactivity. CSF sampling requires a lumbar puncture, which many patients find objectionable. As a result, patterns of A$\beta$, tau, and atrophy are typically identified just at the time when patients are diagnosed with dementia, when irreversible neurodegeneration has occurred and medical intervention has less chance for success.

A potential new biomarker is the disruption of functional brain networks measured by magnetoencephalography (MEG). MEG is non-invasive and low-cost relative to PET scans. It also offers excellent temporal resolution that may capture subtle brain alterations associated with different brain disorders (e.g., AD, epilepsy and post-traumatic stress disorder), a major advantage over current biomarkers \cite{maestu2015multicenter,papadelis2020pediatric, zhang2020classifying}. A further advantage is that MEG captures the fields produced by intraneuronal currents, providing a more direct index of neuronal activity than methods relying on metabolic responses (fMRI, FDG-PET) \cite{hamalainen_magnetoencephalographytheory_1993,khan2019using, baillet_magnetoencephalography_2017}. Electrophysiological measures have already proven useful in tracking brain activity disruption in relation to other known AD pathologies, such as A$\beta$ and tau deposits \cite{canuet_network_2015,stoiljkovic_altered_2018}. Further, synaptic dysfunction (and thus functional disruption) is an important biomarker shown to be more tightly associated with the degree of cognitive impairment than A$\beta$ plaques and tau tangles \cite{blennow2018biomarkers}. Hence, a MEG-based biomarker could provide a unique and more reliable indicator for synaptic dysfunction evaluation during AD progression \cite{yang2019m}. 

To design a MEG-based AD biomarker, there is a need to develop methods that extract quantitative information from MEG brain network data. Prior studies primarily focused on handcrafted, domain-specific (ad-hoc) graph topological properties of brain networks constructed by MEG \cite{pusil2019aberrant}, but also EEG \cite{surya2020complex}, DTI \cite{2018DTI}, and structural MRI data \cite{gomez2018comparison}. However, inductive and automatic network representation learning from raw MEG functional imaging data remains an open problem for accurate AD diagnosis and prognosis. Recently, emerging graph embedding techniques (e.g., deepWalk \cite{deepwalk}, node2vec \cite{node2vec}, Graph2Gauss \cite{g2g}, etc.) enabled automatic learning of hierarchical, heterogeneous and latent network representations from original complex and high-dimensional graphs in irregular domains. Examples include social networks \cite{wang2019user,hu2019sparse}, protein networks \cite{wang2019identification}, and brain networks \cite{rosenthal2018mapping}. The resulting task-independent graph node embeddings can be used as latent features for a variety of machine learning tasks (e.g., link prediction, community detection, node classification, etc.). Incorporating powerful graph representation learning methods in complex human brain network applications could yield a novel and promising tool for characterization of AD progression. 

In order to address the aforementioned problems, we developed a deep learning-based graph Gaussian embedding method for identification and characterization of the early stages of AD using eye-closed resting-state MEG data. First, we constructed subject-specific MEG brain networks based on mutual information in alpha band for brain regions extracted from the Desikan-Killiany atlas~\cite{desikan2006automated}. We focused on the alpha band because it has high signal-to-noise ratio and a growing number of studies report alpha network activity dysregulation in AD \cite{maestu2015multicenter,lopez2014alpha, koelewijn_alzheimers_2017,babiloni_cortical_2014}, but our approach is directly applicable to other frequency bands that have also been implicated in AD \cite{gomez_disturbed_2009,moretti_theta_2015}. We then employed the multiple graph Gaussian embedding model (MG2G)~\cite{xu2019gaussian} to learn highly informative brain network embeddings (patterns) from the original high-dimensional MEG brain networks. Thus, each brain region was represented as a multivariate Gaussian distribution (or Gaussian embeddings) via two vectors, the mean and variance. These latent Gaussian embeddings were subsequently input to multiple classifiers for supervised AD progression prediction. Last, we used the Wasserstein metric \cite{panaretos2019statistical} in the latent space to resolve brain regions with AD-related effects. 

\section*{Methods}
\subsection*{Participants}
MEG data for this study was collected from 76 MCI patients and 53 age-matched neurotypical controls (NC)~\cite{maestu2015multicenter,pusil2019aberrant}. All patients met the MCI core clinical criteria recommended by the National Institute on Aging-Alzheimer Association (NIA-AA) \cite{albert2011}, which are i) concern regarding a change in cognition, ii) impairment in one or more cognitive domains, iii) preservation of independence in functional abilities, and iv) non demented. In addition, MCI patients had signs of neuronal injury, determined by reduced hippocampal volume measured by magnetic resonance imaging, and should thus be considered “MCI due to AD” with an intermediate likelihood~\cite{albert2011}. 

The participant demographic characteristics are shown in Table \ref{tab:dataset_info}. General inclusion criteria were: age between 65 and 80, a modified Hachinski score $\leq4$, a short-form Geriatric Depression Scale score $\leq5$, and T1 magnetic resonance imaging (MRI) within 12 months before the MEG recordings without indication of infection, infarction, or focal lesions (rated by two independent experienced radiologists). Exclusion criteria included a history of psychiatric or neurological disorders other than MCI or AD. Patients were off medications that could affect MEG activity, such as cholinesterase inhibitors, 48 h before recordings.

The MCI patients were followed up for approximately 3 years after their MEG recording session, and were further divided into two groups according to their clinical outcome: the stable MCI group (\textit{sMCI}, $n = 48$) comprising those patients that still fulfilled the diagnosis criteria of MCI at the end of follow-up, and the progressive MCI group (\textit{pMCI}, $n=28$) comprising those patients that met the criteria for probable AD~\cite{mckhann2011diagnosis}.

Patients were recruited from the Hospital Universitario San Carlos (Madrid, Spain), and neurotypical controls were primarily relatives of patients and/or volunteers from the same hospital. All subjects gave a written informed consent and patients received payment for their participation. The study was approved by the local ethics committee (Hospital Universitario San Carlos Ethics Committee, Madrid) and conducted according to the principles of the Declaration of Helsinki.
\begin{table}[!t]
\centering
\caption{\textbf{Detailed demographic characteristics of total participants (Madrid cohort).}}
\renewcommand\arraystretch{1.5}
\begin{tabular}{lllll}
\hline
Item                & NC (n = 53)    & MCI (n = 76) & p value & statistic \\ \hline
Age (years)         & 69.6 $\pm$ 4.6 & 73.7 $\pm$ 5.1 & 0.435 &  $t=0.783$  \\ 
Gender (female, \%) & 72             & 55   & 0.058 &  $\chi^2 =3.58$      \\
Education level   & 3.9 $\pm$ 1.1      & 2.6 $\pm$ 1.3 & 0.936 &  $t=0.081$       \\ 
MMSE score          &  29.4 $\pm$ 0.75    &  26.6 $\pm$ 2.75 & 0.899& $t=0.127$  \\
\hline
\end{tabular}
\label{tab:dataset_info}
\end{table}

\subsection*{MEG acquisition and preprocessing}
Between 3 to 5 minutes of resting-state MEG data were acquired from each participant, depending on subject's cooperation and stillness, while participants were awake with their eyes closed. MEG data was recorded using an Elekta Vectorview system (306-channel probe unit with 204 planar gradiometer sensors and 102 magnetometer sensors) at a sampling rate of 1000 Hz, filtered between 0.1 and 330 Hz. Prior to the MEG recording, a 3D digitizer (Fastrak, Polhemus, Colchester, Vermont, USA) was used to register the locations of 3 anatomical landmarks (right and left preauricular points and the nasion) and 4 head position indicator coils. 

Raw MEG data was pre-processed with the Maxfilter software (Elekta, Stockholm) to perform noise reduction with spatiotemporal filters~\cite{taulu2004suppression, taulu2006spatiotemporal}. We used default parameters (harmonic expansion origin in head frame = [0 0 40] mm; expansion limit for internal multipole base = 8; expansion limit for external multipole base = 3; bad channels automatically excluded from harmonic expansions = 7 s.d. above average; temporal correlation limit = .98; buffer length = 10 s). Intuitively, Maxfilter first applied a spatial filter that separated the signal data from spatial patterns emanating from distant noise sources outside the sensor helmet. It then applied a temporal filter that discarded components of the signal data with time series strongly correlated with the ones from the noise data. The continuous resting-state MEG data was then split into 2 seconds epochs. The epochs were scanned for ocular, muscle, and jump artifacts using the z-score automatic artifact rejection of Fieldtrip software~\cite{oostenveld2011fieldtrip} with the default tutorial parameters. The remaining epochs did not differ (\textit{t}-test, $p = 0.11$) between controls ($\text{M} \pm \text{SD} = 84.8 \pm 7$) and MCI patients ($\text{M} \pm \text{SD} = 81.4 \pm 14.6$). Head movements were monitored during the MEG recordings by continuously activating the head position indicator coils, and not differ (\textit{t}-test, $p = 0.93$) between controls ($\text{M} \pm \text{SD} = 0.84 \pm 0.75$) mm/s and MCI patients ($\text{M} \pm \text{SD} = 0.83 \pm 1.1$) mm/s. Finally, the data were band-pass filtered in the alpha band (8-12 Hz).

\subsection*{Source reconstruction and connectivity analysis}
To reconstruct MEG signals on the cortex, we first computed the forward model using an overlapping spheres model (Huang et al., 1999). MEG cortical maps were then computed using a dynamic statistical parametric mapping approach (dSPM)~\cite{dale2000dynamic}, and time series were derived from 68 anatomical brain regions of interest (ROI) using the Desikan-Killiany atlas~\cite{desikan2006automated} (Fig.~\ref{fig:workflow}AB). Source reconstruction analysis used the Brainstorm software~\cite{tadel2011brainstorm}.

Functional connectivity between every pair of ROIs was assessed similarly to~\cite{maestu2015multicenter} using normalized mutual information, which captures both linear and nonlinear dependencies:
\begin{equation}
    I_{norm}{(X,Y)} = \frac{I(X,Y)}{\sqrt{H(X)H(Y)}}
\end{equation}
where $I(X,Y)$ is the mutual information between ROIs $X$ and $Y$, and $H(X)$ and $H(Y)$ are the corresponding marginal entropies. This yielded a $68\times68$ functional brain connectivity matrix for each participant (Fig.~\ref{fig:workflow}C). Each subject-specific undirected MEG functional brain network (or symmetric functional connectivity matrix) can be described by $G=(V,E)$, $V$ is the set of 68 nodes, and $E$ is the edge set with edge weights equal to the MI value between the corresponding pairs of nodes.

\subsection*{Architecture of MG2G stochastic graph embedding model}
In order to automatically learn multi-scale, nonlinear MEG brain network embeddings (or ``patterns'') in latent space from original high-dimensional MEG brain networks, yet maximally preserving the structure properties for accurate AD progression prediction, we proposed to apply a deep learning-based embedding method called   multiple  graph Gaussian  embedding  model  (MG2G) \cite{xu2019gaussian}. This model is a generalization of the Graph2Gauss architecture \cite{g2g} from single-subject binary graphs to multiple-subject weighted graphs.

\begin{figure*}
\centering
\includegraphics[width=.85\textwidth]{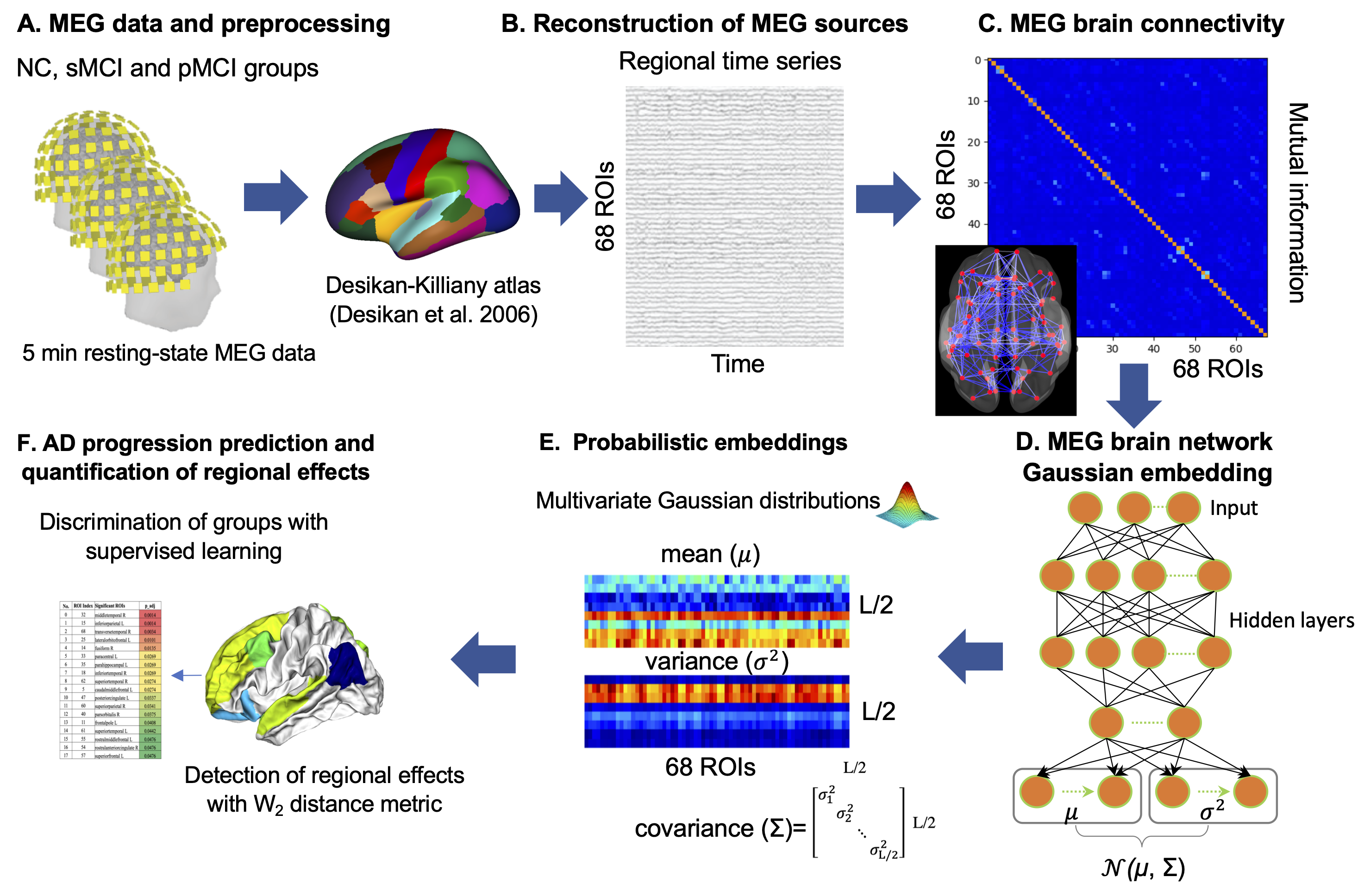}
\caption{Schematic illustration of the graph Gaussian embedding method for predicting Alzheimer’s disease progression using MEG brain networks. (A) 5 min resting-state MEG recordings were obtained for three population groups, neurotypical controls (NC), stable mild cognitive impairment (sMCI) and progressive mild cognitive impairment (pMCI) patients. (B) MEG time courses of 68 brain regions derived from the Desikan-Killiany atlas \cite{desikan2006automated}. (C) Estimation of MEG brain connectivity matrices using mutual information. (D) MG2G model for non-linear stochastic MEG brain network embedding. (E) Stochastic node embedding output in the latent space is a multivariate Gaussian distribution $\mathcal{N}(\mu,\Sigma)$ parameterized by a mean vector ($\mu \in \mathcal{R}^{L/2}$) and variance vector ($\sigma^2 \in \mathcal{R}^{L/2}$), where $L$ refers to the embedding size}. (F) AD progression prediction relied on the MEG Gaussian embedding features to discriminate among population groups. Brain regions with significant AD-related effects were detected using the 2-Wasserstein distance.
\label{fig:workflow}
\end{figure*}

The architecture of the deep learning MG2G model is shown in Fig.~\ref{fig:workflow}D, and detailed information is available in \cite{xu2019gaussian}. Briefly, the MG2G model learns non-linear node embeddings from original high-dimensional brain networks into a stochastic latent space. In the latent space, each node is encoded as Gaussian distributions with two different learned vectors (mean and variance), the mean vector reflects the position of the node in the latent space while the variance provides important uncertainty information. 

Specifically, our model takes the computed undirected and  weighted MEG brain networks $X_p\in \mathbb{R}^{N\times N}, p = 1, 2, ..., P$ ($N$ is the number of brain regions, $P$ is the number of subjects) together as the input. Then, it learns the brain network vector-based encodings using a 3D encoder, which maps the original MEG networks into an intermediate representation through a sequence of hidden layers $h_i^k = ReLU(h_i^{k-1}W_i^k+b^k)$, where $W_i^k \in \mathbb{R}^{N \times D}$ and  $b_i^k \in \mathbb{R}^{D}$ ($k$ denotes the index of hidden layer). Here, due to the learning stability property of G2G model~\cite{g2g}, we used only a single hidden layer ($k = 1$) of size $D = 64$ in our 3D model implementation. The outputs of MG2G are node-wise low-dimensional multivariate Gaussian distributions $P_i = \mathcal{N}(\mu_i, \Sigma_i), i = 1,2,..., N$ parameterized by the mean vector $\mu_i$ and the covariance $\Sigma_i$, where $\mu_i = h_i^kW_{\mu}+ b_{\mu}$, with $W_{\mu} \in \mathbb{R}^{D \times L/2}$, $b_{\mu} \in \mathbb{R}^{L/2}$, and $L$ denoting the embedding size. The covariance matrix $\Sigma_i$ is defined as a square matrix with variance $\sigma_i$ as its diagonal elements, where $\sigma_i = elu(h_i^kW_{\sigma}+ b_{\sigma}))+1$, with $W_{\sigma} \in \mathbb{R}^{D \times L/2}$ and $b_{\sigma} \in \mathbb{R}^{L/2}$. Finally, all of the parameters of our model including weights ($W_i^k, W_{\mu}, W_{\sigma} $) and biases ($b^k, b_{\mu}, b_{\sigma}$) are learned by minimizing the square-exponential loss function $\mathcal{L} = \sum[{\mathbb{E}_{pos}^2+exp^{-\mathbb{E}_{neg})}}]$, where $\mathbb{E}_{pos}$ and $\mathbb{E}_{neg}$ refer to the Kullback–Leibler (KL) divergence energy of the generated node triplets~\cite{xu2019gaussian} involving positive node pairs and negative node pairs, respectively. In particular, we computed the shortest distances between node pairs based on the edge weights and generated different hops. In order to capture high-order proximity, we sampled valid triplet sets based on the obtained hops~\cite{xu2019gaussian}. Lastly, the neural network was optimized by using the Adam algorithm in TensorFlow 1.14.0~\cite{tensorflow2015} with initial learning rate = 1e-3, maximum number of epochs = 1300 and number of hidden units = 64. 

\subsection*{MG2G model optimization with a link prediction task}
Since the MG2G model is an unsupervised learning model, we carried out a ``link prediction'' experiment to identify the optimal embedding dimension $L$ that best preserves the network structure. In addition, link prediction enables us to assess the effectiveness and stability of the MG2G model in representing the original MEG brain networks in a low-dimensional embedding space. Specifically, the total edges/links in the adjacency matrices of the MEG brain networks were split into three subsets: a training set (85\%), a validation set (10\%) and a test set (5\%). The same number of non-edge links were generated and added into the validation and test sets. Then, we first obtained node-wise brain network stochastic embeddings by training the MG2G model with embedding dimensions $L$ equal to 2, 4, 8, 16 and 32, using the same model settings as described in Section D. 

Based on the node-wise Gaussian embedding results, the probability of different links in the validation and test sets could be predicted by computing the KL divergence scores between every node pairs that constituted the sampled links in the validation and test sets. Finally, by evaluating the performances of different link prediction models using different embedding sizes, we obtained the optimal embedding dimension of the original high dimensional MEG brain network data. 

\subsection*{Application of traditional machine learning classifiers on network embeddings to predict AD}
The MG2G model constructs effective network features by mapping high-dimensional MEG resting-state networks into node-wise stochastic MEG brain network embeddings in a latent space. These embeddings can then be used as features in traditional machine learning classifiers to predict early stages of AD. To assess performance irrespective of a particular classifier, we averaged the classification accuracy of ten popular classifiers available in the scikit-learn library \cite{sklearn_api}, namely ``Nearest Neighbors'',``Linear SVM'', ``RBF SVM'', ``Gaussian Process'', ``Decision Tree'', ``Random Forest'', ``Neural Net'', ``AdaBoost'', ``Naive Bayes'', ``QDA''. Each classifier was trained using a 5-fold cross-validation strategy to distinguish between NC, sMCI, and pMCI patients. Classification was 3-class, but also 2-class for the cases  NC vs. sMCI, sMCI vs. pMCI, and NC vs. pMCI.

\subsection*{Detection of brain regions with AD-related effects using 2-Wasserstein distance}
Brain networks (graphs) are an important data representation, but their application in AD progression prediction suffers from their heterogeneity and high dimensionality, as is often the case in graph analytics \cite{cai_comprehensive_2018}. Graph embedding is an effective way to solve this problem. The MG2G model yields highly informative low-dimensional network embeddings (preserving both local and global graph structure properties) in the latent space, and each node is projected to a multivariate Gaussian distribution. One advantage of this representation is that one can 
easily measure the node embedding similarity using the 2-Wasserstein distance ($W_2$), which quantifies distances between Gaussian probability distributions:
\begin{equation}
\begin{split}
\mathrm{W_2(P_i,P_j)^2} & =  W_2(\mathcal N(\mu_{i},\Sigma_{i}),\mathcal N(\mu_{j},\Sigma_{j}))^2 \\
& = \left \|\mu_i-\mu_j \right \|_2^2+\left \|\Sigma_i^{1/2}-\Sigma_j^{1/2} \right \|_F^2
\end{split}
\label{eq:W2distance}
\end{equation}
where $P_i = \mathcal N(\mu_{i},\Sigma_{i})$ and $P_j = \mathcal N(\mu_{j},\Sigma_{j})$ refer to the encoded multivariate Gaussian distributions for nodes $i$ and $j$; and $\mu_{i}$, $\Sigma_{i}$ are the mean and covariance matrix of the node-wise normal distribution, respectively, and \textit{F} denotes the Frobenius norm.

To identify specific brain regions associated with significant AD-related effects, we compared the $W_2$ node distances between patients belonging to different versus same experimental groups. We exemplify this for the NC versus sMCI comparison, but it applies equivalently to other group comparisons. For each brain region separately, we first computed the $W_2$ distances when i) one subject is from the NC group and the other from the sMCI group (``between-group'' distance); and ii) when both subjects are from the same group (``within-group'' distance, NC or sMCI). We then defined the \textit{Group-wise ROI Cluster Index} (GCI) as the averaged $W_2$-distance of the between- minus within-pairs:
\begin{equation}
\begin{split}
GCI = & E_{i\epsilon{NC}, j \epsilon{sMCI}} \{W_2(P_i,P_j)^2\} \\
& - E_{i,j \epsilon{NC}} \{W_2(P_i,P_j)^2\}/2\\
& - E_{i,j \epsilon {sMCI}} \{W_2(P_i,P_j)^2\}/2.
\end{split}
\label{eq:combined_dist}
\end{equation}

A positive GCI for a given brain region suggests network alterations between the NC and sMCI groups. This is because the embeddings would tend to cluster in the same area in the latent space for within-group subjects, but map to distant areas for between-group subjects.

\subsection*{Statistical testing}
We used non-parametric statistical inference that does not make assumptions
on the distributions of the data \cite{maris_nonparametric_2007,pantazis_comparison_2005}. Specifically, to assess the statistical significance of the GCI results for each brain region, we performed permutation tests. Under the null hypothesis of no group effects, we randomly exchanged the labels of the NC and sMCI subjects, each time recomputing a new cluster index, GCI*, where (*) denotes a permutation sample. Repeating the permutation procedure 1000 times yielded an empirical distribution, which enabled us to estimate the p-value of the GCI of the original data. We controlled for multiple comparison over the 68 brain regions using false discovery rate (FDR) at a 0.05 level.

\section{Results}
\subsection*{Optimal embedding dimension of the MEG brain networks}
To optimize the representation of MEG brain networks in the embedded space of the MG2G model, we carried out a link prediction experiment. Link prediction is a common problem in network science that aims to assess the ability to find missing links in a network. To this goal, MEG brain networks were first computed for each subject by estimating the mutual information (MI) in the alpha band between every pair of 68 cortical regions derived from the Desikan-Killiany atlas\cite{desikan2006automated} (Fig.~\ref{fig:workflow}ABC). Then, for the link prediction task, we held out a set of edges/non-edges from the MEG networks before training the MG2G model with different embedding dimensions ($L$=2, 4, 8, 16, 32). 

\begin{figure}[!t]
\centering
\includegraphics[width=\linewidth]{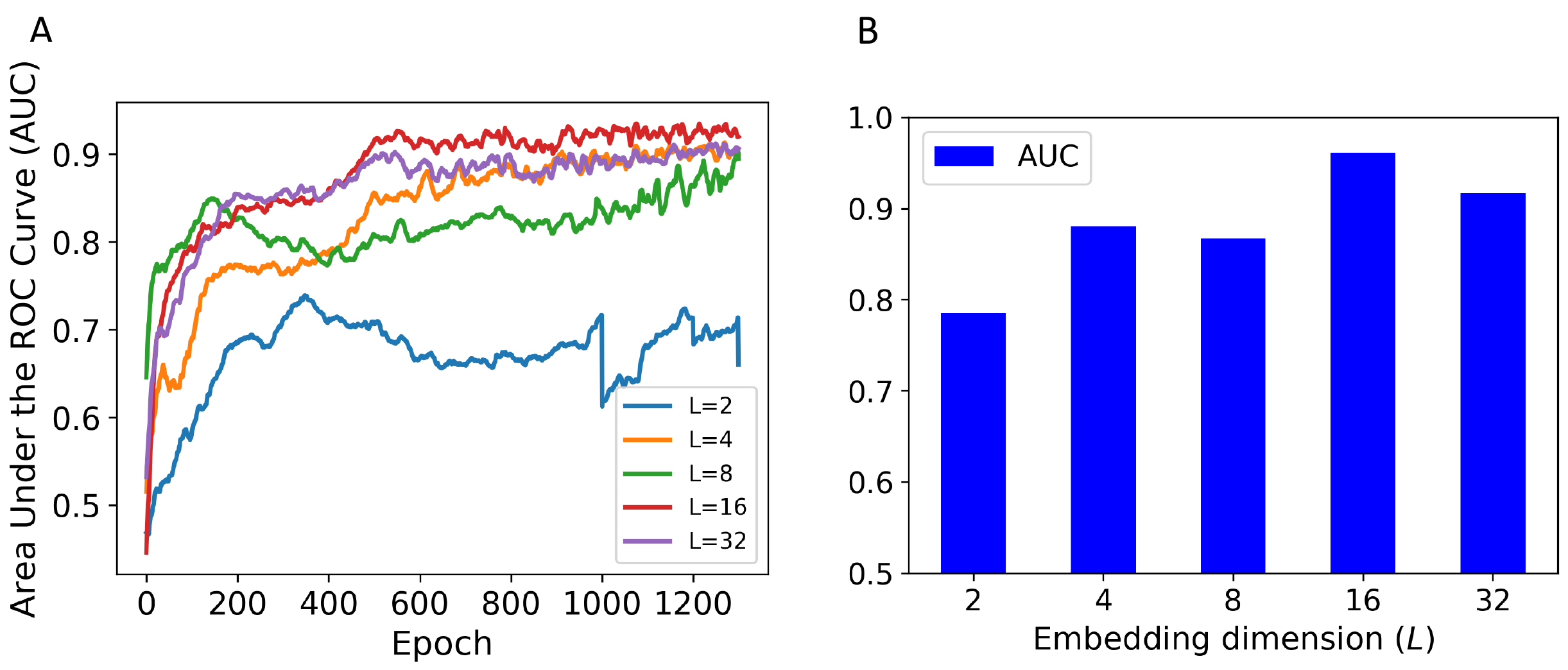}
\caption{Performance of the MG2G brain network embedding model in a link prediction task for different values of embedding size ($L=2,4,8,16,32$). (A) Link prediction AUC results for the validation dataset with increasing number of training epochs. (B) Link prediction AUC results for the test dataset.}
\label{fig:lp_val_test}
\end{figure}

The AUC (area under the ROC curve) results for the validation set are shown in Fig.~\ref{fig:lp_val_test}A. Node embeddings of dimension $L=16$ achieved the best AUC performance. The worst performance was for embeddings of dimension $L=2$ because very low dimensional embeddings cannot sufficiently capture the representational information of the original brain networks. Node embeddings of $L = 32$ probably exceeded the latent dimension of the MEG networks and thus may have included higher levels of noise, which reduced their performance to levels similar to the $L = 4$ case. The AUC results for the test set are shown in Fig.~\ref{fig:lp_val_test}B. Similar to the validation set results, the embedding size $L=16$ resulted in the best performance.

\subsection*{Prediction of AD progression using MEG network embeddings}
The disruption of brain functional connectivity caused by early stages of AD could lead to subtle alterations that are encoded in the MEG network embeddings. We tested this prediction by applying the MG2G model, with embedding size $L$=16, to learn the low-dimensional Gaussian distributed MEG brain network embeddings. We then input the embeddings to multiple traditional machine learning classifiers (``Nearest Neighbors'',``Linear SVM'', ``RBF SVM'', ``Gaussian Process'', ``Decision Tree'', ``Random Forest'', ``Neural Net'', ``AdaBoost'', ``Naive Bayes'', ``QDA'') available in the scikit-learn library \cite{sklearn_api} to distinguish between NC, sMCI, and pMCI patients. To assess how informative are the MG2G embeddings, we also compared the results obtained with the MG2G method against another prevalent point vector-based graph embedding method, the node2vec \cite{node2vec}. We assessed node2vec performance with different embedding sizes ($L$ = 8, 16) and hyperparameters (p and q values), which control the neighborhood exploration in node2vec. 

\begin{figure}[!t]
\centering
\includegraphics[width=\linewidth]{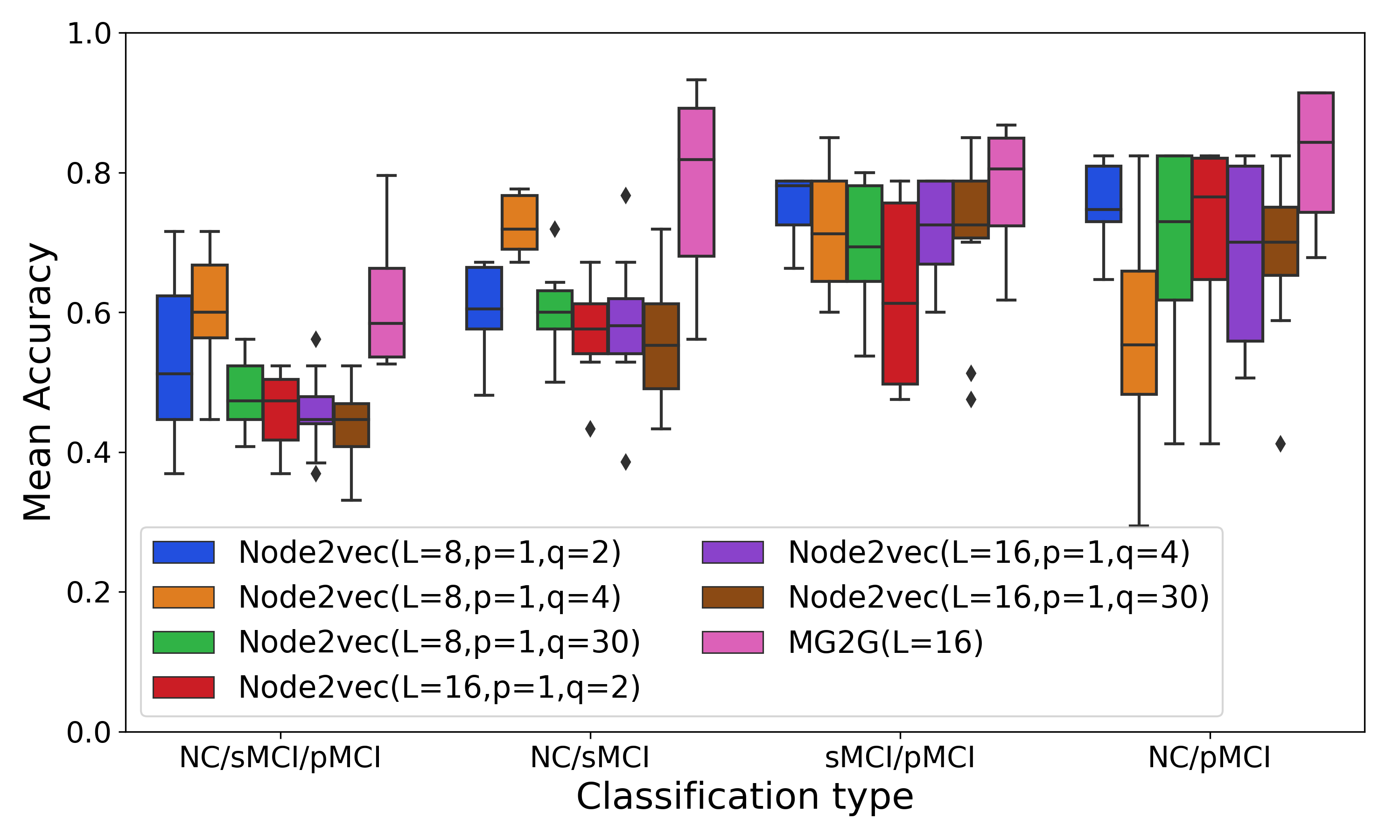}
\caption{Mean accuracy for supervised classification of the NC, sMCI, and pMCI groups. Accuracy is averaged across 10 popular traditional classifiers, when receiving as input the vector-based MEG brain network embeddings obtained from the MG2G and node2vec models. Each individual classifier was trained using a 5-fold cross-validation procedure. Pink color corresponds to the MG2G method with embedding size $L=16$, and the remaining colors correspond to the node2vec method with different node neighbor sampling parameters ($p=1; q=2, 4, 30$) and embedding sizes ($L=8, 16$). Box plots indicate the first and third quartiles, with horizontal lines marking the median, and whiskers the minimum or maximum.}
\label{fig:supervised_learning}
\end{figure}

The mean accuracy across the 10 traditional classifiers for each of the network embedding methods are shown in Fig.~\ref{fig:supervised_learning}. We evaluated performance in both 3-class and 2-class classification tasks. MG2G achieved high performance with 61\% for NC/sMCI/pMCI classification, 79\% for NC/sMCI classification, 78\% for sMCI/pMCI classification, and 82\% for NC/pMCI classification. We note that node2vec had consistently lower performance, with best results achieved when embedding size was $L=8$, and node neighbor sampling parameters $p=1$ and $q=4$. Receiver operating characteristic area-under-the-curve (AUC) plots are shown in Suppl. Figs. 2 and 3. Our method outperformed the baseline models in AUC results for all four different prediction tasks. Overall, the stochastic MG2G graph embedding method yielded highly informative features that learned the latent hierarchical and non-linear MEG brain network patterns in AD early stages.

\subsection*{Brain regions with significant AD-related effects in the MEG network embedding space}

The MG2G model resulted in graph embeddings that represented every brain region (node) by a multivariate Gaussian distribution in a latent space. To pinpoint AD-related MEG network alterations in space, we assessed which brain regions had embeddings that were statistically different for different population groups. This was accomplished by computing, for each node separately, the embedding distance between every pair of patients using the 2-Wasserstein distance ($W_2$). Then a cluster index (GCI), defined as the difference of between-group minus within-group node distances, assessed the extent to which different groups clustered in the latent space. A positive GCI for a given brain region would suggest embeddings that are distant for patients of different experimental groups, but nearby for patients of same experimental groups.

\begin{figure*}
\centering
\includegraphics[width=.9\textwidth]{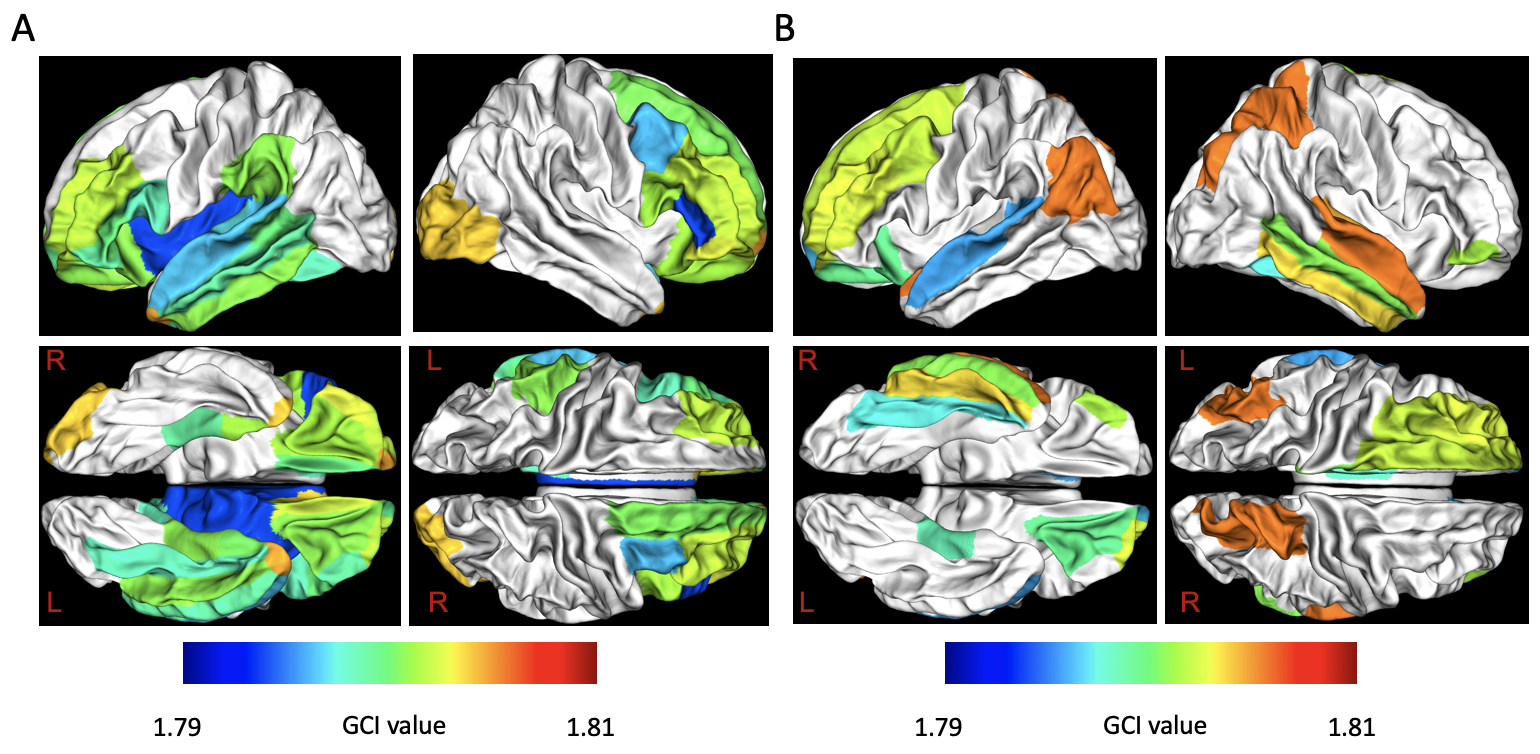}
\caption{Brain regions with significant MEG network alterations due to AD progression. Effects were quantified with a positive GCI ($p<0.05$, one-sided permutation test; FDR-corrected for multiple comparisons). (a) NC vs. sMCI comparison. (b) sMCI vs. pMCI comparison.}
\label{fig:sig_rois}
\end{figure*}

We found 34 brain regions with significantly positive GCI values between the NC and sMCI groups ($p<0.05$, permutation test, FDR corrected for multiple comparisons). The label, lobe name, GCI value, and p-value of each brain region are listed in Table~\ref{tab:significant_ROI_NS}. A large proportion of the significant ROIs concentrated in the frontal and temporal lobes, with only a few in the occipital and parietal lobes. Further, there was a hemispheric asymmetry with 20 ROIs in the left compared to 14 ROIs in the right hemisphere. An equivalent analysis for the sMCI vs. pMCI groups yielded 18 brain regions with significantly positive GCI values, encompassing mainly frontal, temporal and parietal lobes (Table~\ref{tab:significant_ROI_SP}). Of these, 10 regions were in the left hemisphere with the remaining 8 in the right hemisphere. The brain regions detected by the NC vs. sMCI and sMCI vs. pMCI comparisons had only a partial overlap, with seven brain regions common in the two tables: \textit{frontal pole R, parahippocampal L, lateral orbitofrontal L, pars orbitalis R, superior temporal L, rostral middle frontal L, rostral anterior cingulate R}. All brain regions with significant AD-related effects are plotted in Fig.~\ref{fig:sig_rois}. Lastly, a comparison of the NC vs. pMCI groups yielded 19 brain regions, and results are detailed in the supplementary material (Suppl. Table 1 and Suppl. Fig. 1).

\begin{table}[!t]
\centering
\caption{Brain regions with AD-related effects between NC and sMCI}
\begin{tabular}{ccccc}
\hline
No. & ROI label           & Lobe name & $GCI$ value & $p$ value\\\hline
1   & superior temporal L         & Temporal  & 1.790     & 0.028   \\
2   & transverse temporal L       & Temporal  & 1.791     & 0.039   \\
3   & parahippocampal R          & Temporal  & 1.793     & 0.028   \\
4   & fusiform L                 & Temporal  & 1.793     & 0.028   \\
5   & middle temporal L           & Temporal  & 1.794     & 0.028   \\
6   & entorhinal R               & Temporal  & 1.796     & 0.028   \\
7   & entorhinal L               & Temporal  & 1.797     & 0.026   \\
8   & inferior temporal L         & Temporal  & 1.798     & 0.017   \\
9   & parahippocampal L          & Temporal  & 1.798     & 0.025   \\
10  & temporal pole R             & Temporal  & 1.799     & 0.017   \\
11  & temporal pole L             & Temporal  & 1.805     & 0.017   \\
12  & isthmus cingulate L         & Parietal  & 1.793     & 0.028   \\
13  & supramarginal L            & Parietal  & 1.798     & 0.042   \\
14  & lateral occipital R         & Occipital & 1.799     & 0.048   \\
15  & pars triangularis R         & Frontal   & 1.787     & 0.028   \\
16  & caudal middle frontal R      & Frontal   & 1.790     & 0.039   \\
17  & rostral anterior cingulate R & Frontal   & 1.792     & 0.025   \\
18  & pars opercularis L          & Frontal   & 1.793     & 0.039   \\
19  & medial orbitofrontal R      & Frontal   & 1.794     & 0.017   \\
20  & pars orbitalis L            & Frontal   & 1.794     & 0.026   \\
21  & frontal pole L              & Frontal   & 1.795     & 0.017   \\
22  & superior frontal R          & Frontal   & 1.795     & 0.035   \\
23  & lateral orbitofrontal R     & Frontal   & 1.796     & 0.017   \\
24  & pars triangularis L         & Frontal   & 1.796     & 0.039   \\
25  & pars opercularis R          & Frontal   & 1.796     & 0.045   \\
26  & rostral middle frontal R     & Frontal   & 1.797     & 0.035   \\
27  & pars orbitalis R            & Frontal   & 1.798     & 0.017   \\
28  & lateral orbitofrontal L     & Frontal   & 1.798     & 0.025   \\
29  & frontal pole R              & Frontal   & 1.800     & 0.026   \\
30  & rostral middle frontal L     & Frontal   & 1.800     & 0.033   \\
31  & medial orbitofrontal L      & Frontal   & 1.801     & 0.017   \\
32  & rostral anterior cingulate L & Frontal   & 1.803     & 0.026   \\
33  & insula L                   & .         & 1.785     & 0.026   \\
34  & banks sts L                 & .         & 1.796     & 0.028   \\
\hline
\end{tabular}
\label{tab:significant_ROI_NS}
\end{table}

\begin{table}[!t]
\centering
\caption{Brain regions with significant AD-related effects between sMCI and pMCI}
\begin{tabular}{ccccc}
\hline
No. & ROI label           & Lobe name & $GCI$ value & $p$ value\\\hline
1   & middle temporal R           & Temporal  & 1.794    & 0.001 \\
2   & transverse temporal R       & Temporal  & 1.799    & 0.003 \\
3   & fusiform R                 & Temporal  & 1.789    & 0.013 \\
4   & parahippocampal L          & Temporal  & 1.791    & 0.027 \\
5   & inferior temporal R         & Temporal  & 1.798    & 0.027 \\
6   & superior temporal R         & Temporal  & 1.801    & 0.027 \\
7   & superior temporal L         & Temporal  & 1.786    & 0.044 \\
8   & inferior parietal L         & Parietal  & 1.806    & 0.001 \\
9   & posterior cingulate L       & Parietal  & 1.790    & 0.034 \\
10  & superior parietal R         & Parietal  & 1.801    & 0.034 \\
11  & lateral orbitofrontal L     & Frontal   & 1.793    & 0.012 \\
12  & paracentral L              & Frontal   & 1.796    & 0.027 \\
13  & caudal middle frontal L      & Frontal   & 1.799    & 0.027 \\
14  & pars orbitalis R            & Frontal   & 1.795    & 0.037 \\
15  & frontal pole L              & Frontal   & 1.787    & 0.041 \\
16  & rostral middle frontal L     & Frontal   & 1.800    & 0.048 \\
17  & rostral anterior cingulate R & Frontal   & 1.787    & 0.048 \\
18  & superior frontal L          & Frontal   & 1.799    & 0.048\\
\hline
\end{tabular}
\label{tab:significant_ROI_SP}
\end{table}

\section{Discussion}
In this study, we developed a deep learning-based method, called MG2G model, which automatically learns low-dimensional stochastic MEG brain network embeddings from original high-dimensional MEG brain network data. In the latent space, every brain region (ROI) was transformed into a multivariate Gaussian distribution represented by the mean and the variance vectors. We showed that the embeddings in the latent space captured highly informative features that can be used to effectively predict AD progression from NC to sMCI to pMCI patients. Specifically, using the network embeddings, an ensemble of 10 traditional machine classifiers was able to discriminate the three groups with average performance 61\% for NC/sMCI/pMCI classification, 79\% for NC/sMCI classification, 78\% for sMCI/pMCI classification, and 82\% for NC/pMCI classification. Classification results achieved by the MG2G model were higher than the node2vec model \cite{node2vec}, a different deterministic deep-learning model that also learns feature representations from networks. This suggests that the MG2G model can generate more informative MEG network representations in the latent space than the  popular node2vec model, at least in the case of predicting AD progression using MEG data.

 While some graph theoretic measures focus on nodes, such as node degree \cite{RUBINOV20101059}, the majority of traditional brain network analysis methods detect anatomical or functional network alterations either i) at global scale, such as the graph-theoretic measures of  small-worldness, clustering coefficient, and transitivity \cite{RUBINOV20101059}, or ii) at specific network edges \cite{maestu2015multicenter,canuet_network_2015,lopez-sanz_functional_2017}. In contrast, a key advantage of the MG2G model is that network embeddings provide a natural way to identify experimental effects at the level of brain regions (network nodes) rather than brain connections (network edges). Namely, network nodes are represented as vectors in the latent space, hence providing a direct way to quantify node distances, and detect node-specific experimental effects akin to the GCI measure defined in the current study.
 
The MG2G method learns the network embeddings in a completely unsupervised way. Therefore, brain network data are summarized into a low-dimensional representation in an objective way, which can be valuable for subsequent downstream tasks, such as link prediction or clustering. For example, a future study could investigate how brain networks of AD patients cluster in the latent space, which may lead to an automatic detection of AD subtypes. Such investigation would be timely, given that past literature has differentiated at least three AD subtypes (typical, non-limbic, and limbic predominant) based on the distribution of the tau protein \cite{murray_neuropathologically_2011} or patterns of brain atrophy \cite{whitwell_neuroimaging_2012}.

Another advantage of our unsupervised approach is that we can leverage the benefits of deep-learning in brain imaging datasets of small size. In fact, our method is the generalization of Graph2Gauss \cite{g2g}, which learns the embedding of a single graph. In contrast, graph convolutional networks \cite{wang2019spatial}, another family of deep-learning approaches tuned to graphs, are supervised approaches and thus may necessitate much larger datasets to prevent overfitting.   

Using our node-specific analysis in the latent space, we identified that the majority of brain regions with AD-specific effects localized in the frontal and temporal lobes. This includes the parahippocampal cortex, which is associated with visuospatial and episodic memory processing and is consistent with profound memory deficits in AD \cite{aminoff2013role}. The aberrant brain regions identified when contrasting the NC, sMCI, and pMCI groups overlap with previous studies using different imaging modality data, such as FDG-PET \cite{sepulcre2016vivo,sepulcre2018neurogenetic}, MRI \cite{lee2019toward}, and fMRI \cite{hojjati2017predicting}. For example, Lee et al. \cite{lee2019toward} proposed a regional abnormality detection approach for AD diagnosis based on T1 MRI data, and demonstrated that the sensorimotor cortex, temporal lobe and subcortical regions are more closely related to prediction of MCI vs. NC and pMCI vs. sMCI. Additionally, two previous FDG-PET-based AD progression studies \cite{mosconi2004mci,anchisi2005heterogeneity} investigated AD progression biomarkers through evaluating the cerebral glucose metabolism changing patterns, with results indicating that pMCI patients showed reduced cerebral glucose metabolitic rates in some distinct regions (i.e., inferior parietal, posterior cingulate and medial temporal cortices) compared to sMCI patients who did not progress to dementia. The corroborating evidence of regional AD-related effects across multiple biomarkers is consistent with the multifaceted pathology of AD. Different biomarkers have been shown to be sensitive to different pathological aspects of AD and can predict AD conversion. For example, structural MRI and FDG-PET studies have reported sMCI/pMCI classification accuracies ranging from 64\% to 82.5\% \cite{yu_multi-task_2014,the_alzheimers_disease_neuroimaging_initiative_prediction_2016,liu_view-centralized_2015,beheshti_classification_2017}, which is comparable with our MEG findings of 78\% sMCI/pMCI classification accuracy. 

Functional connectivity disruption in AD patients is a well-established finding with most MEG/EEG studies describing a lower functional connectivity in dementia stage, however hyper-synchronization is also typically seen at the MCI stage, which predicts conversion to dementia \cite{pusil_hypersynchronization_2019}. This pattern of hyper/hypo-connectivity is present mainly in higher frequency bands, such as the alpha band investigated here, but also the beta band \cite{gomez2009disturbed, cuesta2015influence}. The analysis of MEG/EEG data at the level of functional brain networks (graphs) is an emerging field in AD. Overall, AD patients show a tendency towards random structure on these networks~\cite{engels_alzheimers_2017}, revealing a significant disorganization with decreased small world, clustering and transitivity, and a more segregated hierarchical source-level brain structure as quantified by increased modularity during resting state \cite{lopez2017network}. In terms of spatial localization of AD pathological effects, while prior findings are not entirely consistent to draw firm conclusions, most AD-related changes are located or involve, parietal and temporal areas \cite{babiloni_abnormal_2004,lopez-sanz_functional_2017,engels_alzheimers_2017}, but also frontal regions \cite{lopez2017network,engels2017directional}, which agree with the areas identified in this study. Since most prior works analyze sensor signals, and not reconstructed source time courses, localization is typically abstracted at the level of brain lobes and not at higher spatial resolution.
 
AD has been commonly found to produce greater atrophy levels in the left compared to the right hemisphere~\cite{donix2013apoe, weise2018left}, and particularly affecting medial temporal lobe structures such as hippocampus or parahippocampal cortex. These previous results seem to agree with those reported here, also showing more AD-related regions in the left hemisphere, with particular differences in this regard over temporal structures in the NC vs. sMCI comparison. Interestingly, some of the regions showing AD-related effects in both comparisons have been commonly reported to be affected in both the early and late stages of the disease and to play a particularly relevant role in network disruption along the progression of this disease, which highlights the clinical relevance of the present results. For instance, a recent work by Yu et al.~\cite{yu2017selective} demonstrated that medial temporal lobe contribution to network organization is disrupted among AD patients, which is in agreement with parahippocampal alterations reported here. Furthermore, anterior cingulate connectivity with other brain regions has been proven highly predictive for the conversion to AD~\cite{lopez2014alpha}, and its functional connectivity patterns seem to be affected even earlier in the course of the disease in older adults with subjective cognitive decline~\cite{lopez2017functional}.

Regarding frontal regions, presenting clear AD-related effects in both groups, previous studies from our group have shown an association of the local oscillatory activity (specifically in the alpha band) and the level of amyloid-$\beta$ accumulation~\cite{nakamura2018electromagnetic}, which support the relevance of alpha band activity over these regions in the context of AD. 

AD is a network-based (connectome) disease. The three earliest biomarkers of AD - tau tangles, amyloid-$\beta$ plaques, and synaptic dysfunction - are not randomly distributed in the brain; rather they have characteristic spatial patterns that appear to follow large-scale brain systems or connectivity networks. Consequently, the trajectory of tau, amyloid-$\beta$, and functional connectivity networks in the early stages of AD could provide a potential disease biomarker with high diagnostic power, enabling early detection and monitoring of disease progression. Thus, a future challenge is to modify our MG2G method to jointly embed multimodal data of functional and protein brain networks.

\section{Conclusion}
Using our MG2G stochastic graph embedding model, we mapped high-dimensional MEG resting-state  networks of MCI patients into node-wise stochastic network embeddings in a low-dimensional latent space. These embeddings were then used as features in traditional  machine learning classifiers to the downstream task of predicting early stages of AD. We  achieved high performance with 61\% for NC/sMCI/pMCI classification, 79\% for NC/sMCI classification, 78\% for sMCI/pMCI classification, and 82\% for NC/pMCI classification. We then identified specific brain regions with MCI-related effects, summarized in Tables ~\ref{tab:significant_ROI_NS} and \ref{tab:significant_ROI_SP}, by exploring group clustering patterns in the latent space derived from the 2-Wasserstein distance measure. Overall, our findings showed that the MG2G unsupervised graph embedding model combined with traditional classifiers is effective in the downstream task of AD prediction with MEG data, without the need of very large data sets that are typically required in supervised deep-learning approaches.

\section*{Acknowledgment}
This work was supported by a J-Clinic for Machine Learning in Health at MIT grant award to D.P.; National Institutes of Health award R01AG052653 to Q.L.; Spanish Ministry of Economy and Competitiveness grant awards PSI2009-14415C03-01 and PSI2012-38375-C03-01 and Madrid Neurocenter grant award B2017/BMD-3760 to F.M.; and the Spanish Ministry of Science, Innovation and Universities, Juan de la Cierva-Formación grant award FJC2018-037401-I to D.L.S..


\end{document}